\documentclass[conference,a4paper]{IEEEtran}

\normalsize

\usepackage{graphicx,epsfig,epstopdf,amsmath,amssymb,enumerate} 
\usepackage{cite}  
\usepackage{algpseudocode}
\usepackage{amsthm}

\newtheorem*{theorem*}{Theorem}
\newtheorem*{lemma*}{Lemma} 

\ifCLASSINFOpdf
\else
\fi

\usepackage[caption=false,font=footnotesize]{subfig}
\begin{document}
\topskip 16pt 

\title{Genetic Algorithm Assisted Hybrid Beamforming for Wireless Fronthaul}

%\author{Shangbin~Wu, xxx,~\IEEEmembership{Member,~IEEE,}  xxx,~\IEEEmembership{Senior Member,~IEEE,} xxx,~\IEEEmembership{Senior Member,~IEEE,} and xxx,~\IEEEmembership{Fellow,~IEEE} 
\author{Shangbin Wu\\
      \IEEEauthorblockA{\IEEEauthorrefmark{1}% 1st affiliations
Samsung R\&D Institute UK, Communications House, Staines-upon-Thames, TW18 4QE, United Kingdom\\ shangbin.wu@samsung.com}  
%\thanks{The authors gratefully acknowledge the support of ....}% <-this % stops a space   
    
%\thanks{S. Wu is with Samsung R\&D Institute UK, Communications House, South Street, Staines-upon-Thames, TW18 4QE, United Kingdom (E-mail: shangbin.wu@samsung.com).}% <-this % stops a space
%\thanks{el-H. M. Aggoune and M. M. Alwakeel are with the Sensor Networks and Cellular Systems (SNCS) Research Center, University of Tabuk, P. O. Box: 6592-2, 47315/4031 Tabuk, Saudi Arabia (E-mail: \{haggoune.sncs, alwakeel\}@ut.edu.sa).}
%\thanks{X. -H. You is with the Mobile Communications Research Laboratory, Southeast
%University, Nanjing, 211189, China (E-mail: xhyu@seu.edu.cn).}% <-this % stops a space
%\thanks{TCOM version based on Michael Shell's bare{\textunderscore}jrnl.tex version 1.3.}
}

\maketitle
\begin{abstract}
This paper proposes a genetic algorithm assisted hybrid signal to leakage plus noise ratio (SLNR) beamforming design for wireless fronthaul scenario. The digital precoder of the proposed hybrid SLNR beamforming is expressed in closed-form. Highly limited phase resolution (one-bit resolution) is assumed at the phase shifters at the analog precoder. The analog precoders maximizing the approximated sum rate are presented. Genetic algorithms are used to search for optimal solutions of one-bit analog precoders. In contrast to common assumptions on perfect knowledge of the \textit{true} channel matrix at the transmitter, the proposed method relies only on the distorted  channel matrix after the analog precoder. Performance of the proposed hybrid SLNR beamforming with limited phase resolution at the analog precoder can achieve performance close to digital beamforming in single cell wireless fronthaul scenarios. It is also shown that hybrid beamforming can result in undesired beams causing intercell interference in multicell wireless fronthaul scenarios.
\\
\\
{\it \textbf{Keywords}} -- Wireless fronthaul, hybrid beamforming, SLNR, genetic algorithm.

\end{abstract}

\IEEEpeerreviewmaketitle

%%%%%%%%%%%%%%%%%%%%%%%%%%%%%%%%%%%%%%%%%%%%%%%%%%%%%%%%%%%%%%%%%%%%%%%%%%%%%%%%%%%%%%%%%%%%
%%%%%%%%%%%%%%%%%%%%%%%%%%%%%%%%%%%%%%%%%%%%%%%%%%%%%%%%%%%%%%%%%%%%%%%%%%%%%%%%%%%%%%%%%%%%

\section{Introduction}
\begin{figure}[t]
\centering
\includegraphics[width=3.5in]{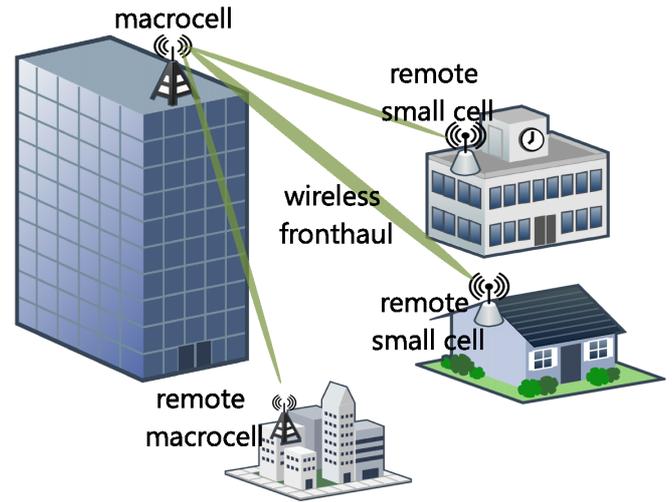}    
\caption{Diagram of wireless fronthaul.}
\label{fig_Wireless_fronthaul}
\end{figure}

The emerging concept of cloud radio access network (CRAN) forms a network by centralizing baseband units (BBUs) in a data center while distributing radio remote heads (RRHs) across the whole network layout. The common public radio interface (CPRI) has been used to connect the BBU and RRUs. However, the fibre implementation of CPRI introduces high cost in fronthaul and limits flexibility of the fronthaul network. Nowadays, the wide spectrum in millimeter wave (mmWave) frequency bands provides the potential of fronthaul links with quasi-fibre throughput and enhanced flexibilities. A working group has been established in IEEE to standardize the next generation fronthaul interface (NGFI) \cite{NGFI} which includes the discussion of wireless fronthaul \cite{CMCC}. A typical structure of wireless fronthaul is depicted in Fig. \ref{fig_Wireless_fronthaul}, where the macrocell can be regarded as the data center and is connecting remote nodes wirelessly.

To support high throughput wireless fronthaul links, beamforming techniques are compulsory. Recently, hybrid beamforming has widely attracted attentions from both academia and industry. It is expected to be used in the fifth generation (5G) millimeter wave systems, where multiple-input multiple-output (MIMO) systems are equipped with a massive number of antennas and a limited number of radio frequency (RF) chains to compensate severe pathloss. A hybrid beamforming system consists of a digital precoder and an analog precoder in a transceiver. In practice, the two precoders can further split into two units, i.e., digital precoder at the BBU and analog precoder at the RRH. The design of hybrid beamforming algorithms involves the design of digital and analog precoders. 

There are extensive digital precoders for multiuser scenarios reported in the literature such as zero forcing (ZF) and minimum mean squared error (MMSE) beamforming. One widely used technique is called the block--diagonalized (BD) ZF beamforming, where the digital precoding vector of each user lies in the intersection of null spaces of other users. The complexity of BD ZF beamforming is relatively low. However, BD ZF beamforming must satisfy the dimension condition \cite{Sadek07}, i.e., the number of transmit antennas has to be larger than the total number of receive antennas of interfering users. Later, digital signal to leakage plus noise ratio (SLNR) beamforming was proposed in \cite{Sadek07}. The aim of SLNR beamforming is to suppress the leakage to other users. A key benefit of SLNR beamforming over ZF beamforming is that the dimension condition is not required in SLNR beamforming.

An analog precoder is comprised of a set of phase shifters, which are not able to adjust signal amplitudes. This constant modulus feature makes the design of analog precoder difficult as the optimization problem will become nonconvex. As a result, iterative procedure is usually used to compute analog precoders. Iterative algorithms for analog precoders to maximize mutual information were proposed in \cite{Pi12}. Authors in \cite{Dai15} considered the sub-connected structure and sequentially computed the optimal precoder for each subarray. A joint analog precoder and combiner design for both fully connected structure and sub-connected structures was proposed in \cite{Li17}. Hadamard matrix and Fourier matrix were considered in \cite{Seleem17} as the codebook for the analog precoder. Also, limited phase resolution was assumed in the analog precoder. Orthogonal matching pursuit was used for the hybrid precoder to approximate the optimum precoder. Large system analysis  was used in \cite{Sohrabi17} to assist the design of hybrid beamforming for millimeter wave systems. This design approach was first used for single user scenario and then generalized to multi-user scenario. The design in \cite{Rusu16} and \cite{Ayach14} provided low-complexity solutions to find analog and digital precoders based on the unconstrained optimum precoders. Iterative analog precoder design with quantization was proposed in \cite{Ayach14}, where the phase of each phase shifter can only be selected from a finite set of values. 

The hybrid precoder designs in \cite{Pi12}--\hspace{-0.01cm}\cite{Ayach14} were based on the assumption that certain perfect knowledge of the \textit{true} channel was known to the transmitter, such as the \textit{true} channel matrix, the covariance matrix of the \textit{true} channel, and the unconstrained optimum precoder of the \textit{true} channel matrix. However, these assumptions are less practical in the context of hybrid beamforming and may contradict to the application of analog precoders. The central processing unit (CPU) can only observe the channel after the analog precoder, which has been distorted due to the limited capability of the analog precoder. Then, the CPU will attempt to optimize the analog precoder based on these distorted channel matrices.

%However, hybrid SLNR beamforming has not been studied in the literature. When the number of radio frequency (RF) chains does not satisfy the dimension condition, hybrid SLNR beamforming will significantly outperform hybrid ZF beamforming as we will show later. The contributions of this paper are two--fold. First, hybrid SLNR beamforming for multiuser transmission is proposed. The design principle of the analog precoder for hybrid SLNR is maximize the approximated sum rate. Second, solutions for analog precoders with finite quantization bits are found via genetic algorithms. The performance of hybrid SLNR beamforming is compared with other digital and hybrid beamforming techniques. 

%The design of analog precoders has been extensively studied in the literature. 

The contributions of this paper are listed as follows.
\begin{enumerate}
\item To consider more practical scenarios, the optimization of analog precoder in this paper does not need the perfect knowledge of the \textit{true} channel matrix and is performed using the genetic algorithm.

\item Solutions for analog precoders with one quantization bit are found via genetic algorithms. The proposed implementation can significantly reduce the complexity of beamformed wireless fronthauls.

\end{enumerate} 
 
The remainder of this paper is organized as follows. Section~\ref{sec_System_Model} gives the general description of the system model of hybrid SLNR beamforming as well as design principles of both digital and analog precoders. Section~\ref{sec_GA} presents the genetic algorithm for finding solutions to analog precoders. Section \ref{sec_Results_and_Analysis} provides simulation results and discussions. Conclusions are drawn in Section~\ref{conclusion_section}.
%%%%%%%%%%%%%%%%%%%%%%%%%%%%%%%%%%%%%%%%%%%%%%%%%%%%%%%%%%%%%%%%%%%%%%%%%%%%%%%%%%%%%%%%%%%%
%%%%%%%%%%%%%%%%%%%%%%%%%%%%%%%%%%%%%%%%%%%%%%%%%%%%%%%%%%%%%%%%%%%%%%%%%%%%%%%%%%%%%%%%%%%%
\section{System Model} \label{sec_System_Model}

Let us consider a wireless fronthaul system in time division duplex (TDD) mode using hybrid beamforming system with a macrocell (transmitter) and $K$ remote nodes (receivers). The transmitter is equipped with $N_\mathrm{T}$ transmit antennas and $N_\mathrm{RF}$ RF chains. The number of RF chains is assumed to be no less than the number of remote nodes to support enough data streams, i.e., $N_\mathrm{RF}\geq K$. Let $\mathbf{H}_l$ be the $N_{\mathrm{R},l}\times N_\mathrm{T}$ channel matrix between the $l$th remote node and the transmitter. Denote the transmitted symbol for the $k$th remote node as $s_k$. The received vector $\mathbf{y}_l$ of the $l$th remote node is given by
\begin{align}
\mathbf{y}_l&=\mathbf{H}_l\mathbf{A}\sum\limits_{k=1}^K  \mathbf{D}_k {s}_k+\mathbf{n}_l=\mathbf{H}^{\mathrm{E}}_l\sum\limits_{k=1}^K  \mathbf{D}_k {s}_k+\mathbf{n}_l
\end{align}
where $\mathbf{D}_k$ is a $N_{\mathrm{RF}}\times 1$ digital precoding vector, $\mathbf{n}_l$ is the Gaussian noise vector, $\mathbf{A}$ is a $N_\mathrm{T}\times N_{\mathrm{RF}}$ analog precoding matrix, and $\mathbf{H}^{\mathrm{E}}_l=\mathbf{H}_l\mathbf{A}$ is the effective channel matrix observed at the CPU. In practical systems, further constraints are required by analog precoding matrices. An analog precoding matrix consists of phase shifters only. Also, the value of the phase of a phase shifter is restricted by $B$ bits. The set $\mathcal{S}$ of all analog precoding matrices can be expressed as
%\begin{align}
%\mathcal{S}=&\left\lbrace \mathbf{A} \in  \mathbb{C}^{N_\mathrm{T}\times N_{\mathrm{RF}}}| \right. \nonumber\\
%& \left. \arg a_{mn}\in \{ \frac{2\pi}{2^B},\frac{2\cdot2\pi}{2^B},\cdots, 2\pi \}, |a_{mn}|=\frac{1}{\sqrt{N_\mathrm{RF}}} \right\rbrace
%\end{align}
\begin{align}
\mathcal{S}=&\left\lbrace \mathbf{A} \in  \mathbb{C}^{N_\mathrm{T}\times N_{\mathrm{RF}}}|  \varphi \in \{ \frac{2b\pi}{2^{B}} \}_{b=1}^{2^{B}}, a_{mn}=e^{j\varphi} \right\rbrace
\end{align}
where $a_{mn}$ is the element in the $m$th row and $n$th column in the analog precoding matrix. Although the design principle to be described in following paragraphs can be applied to phase shifters with arbitrary phase resolution, this paper focuses on one-bit phase resolution, i.e., $B=1$, as this can largely reduce the cost of antenna deployments. With the above settings, the SINR of the $l$th remote node in a hybrid beamforming system can be calculated as
\begin{align}
\mathrm{SINR}_l=\frac{\|\mathbf{H}_l\mathbf{A} \mathbf{D}_l \|^2}{M_l\sigma^2+\sum\limits_{k=1,k\neq l}^{K}\|\mathbf{H}_l\mathbf{A} \mathbf{D}_k \|^2}.
\label{equ_SINR}
\end{align}
The aggregate sum rate $R$ of all remote nodes is presented as
\begin{align}
R=\sum\limits_{k=1}^{K}\log_2\left(1+\mathrm{SINR}_k\right).
\label{equ_sumrate}
\end{align}
It is desired to maximize the sum rate when designing the analog and digital precoders. Maximum sum rate is reached when the SINR of each remote node is maximized. However, computing the digital precoder with maximized SINR is a nonconvex problem. Alternatively, \cite{Sadek07} proposed a fully digital method to maximize SLNR. For hybrid beamforming, analog and digital precoder design have not been studied. Therefore, in the following sections, precoder design for hybrid SLNR beamforming will be presented.

\subsection{Digital precoder design}\label{Digital_precoder_design}
The SLNR using hybrid SLNR precoding is expressed as
\begin{align}
\mathrm{SLNR}_l=\frac{\|\mathbf{H}_l\mathbf{A} \mathbf{D}_l \|^2}{M_l\sigma^2+\sum\limits_{k=1,k\neq l}^{K}\|\mathbf{H}_k\mathbf{A} \mathbf{D}_l \|^2}
\end{align}
It can be seen that the product $\mathbf{A} \mathbf{D}_l$ of the analog precoder and the digital precoder can be regarded as an effective precoder. However, it should be noticed that this effective precoder is significantly different from \cite{Sadek07} because of the restrictions on the structure of the analog precoder $\mathbf{A}$. Let $\tilde{\mathbf{H}}^{\mathrm{E}}_l=\left[\mathbf{H}^{\mathrm{E}}_1 \mathbf{H}^{\mathrm{E}}_2\cdots\mathbf{H}^{\mathrm{E}}_{l-1}\mathbf{H}^{\mathrm{E}}_{l+1}\cdots \mathbf{H}^{\mathrm{E}}_K\right]^{\mathrm{T}}$, $\mathrm{SLNR}_l$ can be rewritten as
\begin{align}
\mathrm{SLNR}_l&=\frac{\mathbf{D}_l^{\mathrm{H}}\left(\mathbf{H}^{\mathrm{E}}_l \right)  ^{\mathrm{H}}\mathbf{H}^{\mathrm{E}}_l \mathbf{D}_l}{M_l\sigma^2+\|\tilde{\mathbf{H}}^{\mathrm{E}}_l \mathbf{D}_l \|^2}\nonumber\\
&=\frac{\mathbf{D}_l^{\mathrm{H}}\left(\mathbf{H}^{\mathrm{E}}_l \right)  ^{\mathrm{H}}\mathbf{H}^{\mathrm{E}}_l \mathbf{D}_l}{\mathbf{D}_l^\mathrm{H}\left( M_l\sigma^2\mathbf{I}+\left(\tilde{\mathbf{H}}^{\mathrm{E}}_l \right) ^\mathrm{H}\tilde{\mathbf{H}}^{\mathrm{E}}_l\right) \mathbf{D}_l}.
\label{equ_SLNR_l}
\end{align}
According to \cite{Sadek07}, maximizing (\ref{equ_SLNR_l}) can be solved via the generalized Rayleigh quotient and the optimum digital precoder is expressed as $\mathbf{D}_l\propto \mathbf{v}_1$,
%\begin{align}
%\mathbf{D}_l\propto \mathbf{v}_1 \label{equ_digital_precoder}
%\end{align}
where $\mathbf{D}_l$ needs to be normalized such that the aggregate precoding matrix has norm one, i.e.,  $\|\mathbf{A}\mathbf{D}_l\|=1$ and $\mathbf{v}_1$ is the eigen vector corresponding to the maximum eigen value $\lambda_{\max}^{(l)}$ of 
\begin{align}
\left( M_l\sigma^2\mathbf{I}+\left(\tilde{\mathbf{H}}^{\mathrm{E}}_l \right) ^\mathrm{H}\tilde{\mathbf{H}}^{\mathrm{E}}_l \right)^{-1}\left(\mathbf{H}^{\mathrm{E}}_l \right)  ^{\mathrm{H}}\mathbf{H}^{\mathrm{E}}_l
\label{equ_targe_matrix}
\end{align}
%
%\begin{align}
%\left( M_1\sigma^2\mathbf{I}+\mathbf{A}^\mathrm{H}\tilde{\mathbf{H}}_1^\mathrm{H}\tilde{\mathbf{H}}_1\mathbf{A}\right)^{-1}\mathbf{A}^{\mathrm{H}}\mathbf{H}_1^{\mathrm{H}}\mathbf{H}_1\mathbf{A}
%\label{equ_targe_matrix}
%\end{align}
%
%
%
%\begin{align}
%\left( M_K\sigma^2\mathbf{I}+\mathbf{A}^\mathrm{H}\tilde{\mathbf{H}}_K^\mathrm{H}\tilde{\mathbf{H}}_K\mathbf{A}\right)^{-1}\mathbf{A}^{\mathrm{H}}\mathbf{H}_K^{\mathrm{H}}\mathbf{H}_K\mathbf{A}
%\label{equ_targe_matrix}
%\end{align}
where the superscript $(l)$ denotes the $l$th remote node. Consequently, the resulting $\mathrm{SLNR}_l$ equals $\lambda_{\max}^{(l)}$. 
%$\lambda_{\max}^{(1)}$.
% $\lambda_{\max}^{(K)}$. $\otimes$
\subsection{Analog precoder design}\label{Analog_precoder_design}
Since there is not a direct mapping from SLNR to sum rate, low signal to noise ratio (SNR) approximation is used to design the analog precoder. When the noise power is significantly larger than the power of leakage or interference, i.e., $M_l\sigma_l^2>>\sum\limits_{k=1,k\neq l}^{K}\|\mathbf{H}_k\mathbf{A} \mathbf{D}_l \|^2$, the sum rate, which is also defined as the cost function $f\left(\mathbf{A}\right)$, can be approximated by 
\begin{align}
f\left(\mathbf{A}\right)=R\approx\sum\limits_{k=1}^{K}\log_2\left(1+\lambda_{\max}^{(k)}\right).
\label{equ_sumrate_approx}
\end{align} 
As a result, to maximize the sum rate, the optimum analog precoder is expressed as $\mathbf{A}^{\mathrm{opt}}=\arg\max_{\mathbf{A}\in \mathcal{S}} f\left(\mathbf{A} \right)$. This optimization problem is a combinatorial optimization problem. The total number of possible combinations equals the cardinality of the set $\mathcal{S}$. However, the cardinality of the set $\mathcal{S}$ can be computed as $2^{N_\mathrm{T}N_{\mathrm{RF}}B}$, which is massive for exhaustive search. Therefore, a practical search algorithm for (sub-)optimal solutions is needed.

%\begin{align}
%\mathbf{A}^{\mathrm{opt}}=\arg\max_{\mathbf{A}\in \mathcal{S}} \prod\limits_{k=1}^K (1+\lambda_{\max}^{(k)}).
%\label{equ_A_opt}
%\end{align}

%\begin{align}
%f\left(\mathbf{A}\right)=\max_{k} \left\lbrace \frac{L/K}{\log_2\left(1+\lambda_{\max}^{(k)} \right)}+\tau_k \right\rbrace
%%} \right\rbrace\prod\limits_{k=1}^K (}).
%\end{align}

%\begin{align}
%&\mathbf{A}^{\mathrm{opt}}=\nonumber\\
%&\arg\max_{\mathbf{A}\in \mathcal{S}} \lambda_{\max}\left\lbrace \left( M_l\sigma_l^2\mathbf{I}+\mathbf{A}^\mathrm{H}\tilde{\mathbf{H}}_l^\mathrm{H}\tilde{\mathbf{H}}_l\mathbf{A}\right)^{-1}\mathbf{A}^{\mathrm{H}}\mathbf{H}_l^{\mathrm{H}}\mathbf{H}_l\mathbf{A} \right\rbrace.
%\label{equ_A_opt}
%\end{align}

\subsection{Note}
The \textit{true} channel matrix $\mathbf{H}_l$ cannot be recovered from the distorted channel matrix $\mathbf{H}_l\mathbf{A}$ even if the CPU knows the analog precoder $\mathbf{A}$, because $\mathbf{A}$ is not a square matrix in general. Techniques can be used to estimate the \textit{true} channel matrix but the estimation error was ignored in the literature. On the contrary, the method proposed in this paper depends only on the distorted channel matrix after the analog precoder and not on the  \textit{true} channel matrix. The fitness function only depends on the observed distorted channel and the \textit{true} channel matrix is not needed. Therefore, this is more practical in hybrid beamforming system design.

%it can be seen in Sections \ref{Digital_precoder_design} and \ref{Analog_precoder_design} that neither the digital precoder nor the analog precoder depends on the  \textit{true} channel matrix. These are more practical in hybrid beamforming system design.

\section{Genetic Algorithm for Analogy Precoder Design}\label{sec_GA}
Genetic algorithm \cite{Melanie99} has been widely used to solve combinatorial optimization problem. It mimics the biological evolution by introducing random variations to the chromosomes in the population. Chromosomes with higher fitness in the population will have higher chances of survival. After a few generations, the remained chromosomes in the population are likely to have high fitness values and therefore represent (sub-)optimal solutions. The three basic operations in genetic algorithm, i.e., selection, crossover, and mutation, will be performed iteratively and introduced in later sections. Authors in \cite{Hong14} used a floating-point genetic algorithm to design hybrid beamforming to maximize SINR. However, neither the analog precoder nor the digital precoder has closed-form solutions in \cite{Hong14}. It should be noticed that the hybrid SLNR beamforming proposed in this paper has closed-form solutions to the digital precoder, which largely reduces the computation complexity compared to \cite{Hong14}.

Since the analog precoder consists of $B$-bit phase shifters only, the phase of an element in the analog precoder can be encoded as a $B$-bit chromosome. After binary encoding mechanism is determined, the search for analog precoder with genetic algorithm can start.

%There is a bijection between a phase matrix of the analog precoder and a chromosome. The mapping from a phase matrix to a chromosome is shown in Fig. \ref{fig_Phase2Chromosome}. In the phase matrix in Fig. \ref{fig_Phase2Chromosome}, $b^{q}_{ij}$ represents the $q$th bit of the entry in the $i$th row and $j$th column in the phase matrix. After binary encoding mechanism is determined, the search for analog precoder with genetic algorithm can start.

%\begin{figure}[t]
%\centering
%\includegraphics[width=3.5in]{Phase2Chromosome.eps}    
%\caption{Mapping from a phase matrix of the analog precoder to a chromosome.}
%\label{fig_Phase2Chromosome}
%\end{figure} 

%\subsection{Initialization and Fitness function}
Initialization is realized by randomly choosing $N_\mathrm{P}$ points in the solution space $\mathcal{S}$, where $N_\mathrm{P}$ is the size of the population. A typical value for $N_\mathrm{P}$ is $50$. The population is denoted as $\mathcal{S}_P$. In this paper, each bit of a chromosome is initialized by randomly selected between $0$ and $1$ with equal probability. Also, initialization includes setting the maximum number of generations $N_{\mathrm{gen\_ max}}$ as the condition to exit the iteration process. The value of $N_{\mathrm{gen\_ max}}$ is a tradeoff between complexity and performance. In this paper, as shown in later paragraphs, $150\leqslant N_{\mathrm{gen\_ max}}\leqslant200$ will be reasonable numbers.
%\subsection{Fitness}
In the context of genetic algorithm, the fitness is defined as the probability that a chromosome is able to reproduce. It is straightforward to reuse the approximated sum rate in (\ref{equ_sumrate_approx}) as the fitness function.
%According to (\ref{equ_A_opt}), fitness $f(\mathbf{A})$ of a chromosome $\mathbf{A}$ for analog precoder design can be defined as
%\begin{align}
%f(\mathbf{A})= \prod\limits_{k=1}^K (1+\lambda_{\max}^{(k)})
%\end{align}
%Since the user index $l$ does not affect the description, the subscript $l$ is dropped for compact interpretation. 
A chromosome with larger fitness corresponds to a better solution for the analog precoder and higher SLNR. Also, it  has a higher chance to be selected for crossover, mutation, and survive in the population.

%\subsection{Selection, Crossover, and Mutation}
The selection process is based on the roulette wheel selection \cite{Melanie99}. The selection probability of $\mathbf{A}$ is ${f(\mathbf{A})}/{(\sum\limits_{\mathbf{A}'\in S_\mathrm{P}}f(\mathbf{A}'))}$. Each selection operation randomly picks two chromosomes in the population according to their selection probabilities. The picked chromosomes are called parents. Selection is performed with replacement, i.e., the selected chromosomes will be placed back to the population. Other selection schemes such as Boltzmann selection rank selection can also be found in \cite{Melanie99}.

%\subsection{Crossover and Mutation}
Crossover randomly interchanges bits between the selected pair of chromosomes (parents) according to a predefined crossover probability $p_\mathrm{c}$. A typical value of $p_\mathrm{c}$ is $0.7$ \cite{Melanie99}. The resulting chromosomes are known as children. Next, mutation is performed on the two children by mutating each bit with mutation probability $p_\mathrm{m}$. A typical value of $p_\mathrm{m}$ is $0.001$ \cite{Melanie99}. Then, the mutated children will be placed into the new population.

%\subsection{Iterations}

%\subsection{Iteration}
%The population will be replaced by the new population and the fitness evaluation, selection, crossover, and mutation will be performed iteratively, until the number of generations reaches $N_{\mathrm{gen\_ max}}$. The whole process of genetic algorithm for hybrid SLNR analog precoder design is illustrated in Fig. \ref{fig_Flowchart}.
%\begin{figure}[t]
%\centering
%\includegraphics[width=3.5in]{Flowchart.eps}    
%\caption{Flowchart of analog precoder design for hybrid SLNR beamforming with genetic algorithm.}
%\label{fig_Flowchart}
%\end{figure}

As described in Fig. \ref{fig_algorithm}, the analog precoder is computed using the genetic algorithm in an iterative manner. During iterations, it is assumed that the channel is quasi-static and the remote nodes are sending reference signals to the transmitter. The quasi-static assumption holds in wireless fronthaul use cases as the macrocell and remote nodes are relatively stable comparing to regular mobile communications. In each iteration , the CPU observes the distorted channel matrix $\mathbf{H}_l\mathbf{A}$ without the need of the \textit{true} channel matrix. Relying on channel reciprocity in the TDD mode, the transmitter can adjust the phases of the analog precoder and evaluate the fitness after each adjustment according to the genetic algorithm. The adjustment of analog precoder can either be placed in Line 14 or right after Line 12 in Fig. \ref{fig_algorithm} depending on implementation. 
 
\begin{figure}
\hrulefill
\begin{algorithmic}[1]
%\State Measure the real part of noise plus interference samples $\left\lbrace \tau_a \right\rbrace_{a=1}^{N_s}$
\State Initialize population
\State Generation number $g$=1;
\While {($g\leqslant N_{\mathrm{gen\_max}}$)} 
\State Remote notes transmit reference signals to the macrocell;
\For {$\beta=1,2,...,N_{\mathrm{P}}$}
\State Adjust analog precoder $\mathbf{A}$ according to chromosome $\beta$ in the population;
\State Measure distorted effective channel $\mathbf{H}_k\mathbf{A}, \forall k$;
\State Calculate fitness $f\left( \mathbf{A_k}\right)$;

\EndFor
\State Perform selection, crossover, and mutation;
\State Update population;

\State $g=g+1$;

\EndWhile
\State Output the chromosome with the highest fitness in the population as the analog precoder;

\end{algorithmic}
\hrulefill
\caption{Procedure of analog precoder computation with genetic algorithm.}
\label{fig_algorithm}
\end{figure}

\section{Results and Analysis}\label{sec_Results_and_Analysis}
\begin{figure}[t]
\centering
\includegraphics[width=3.5in]{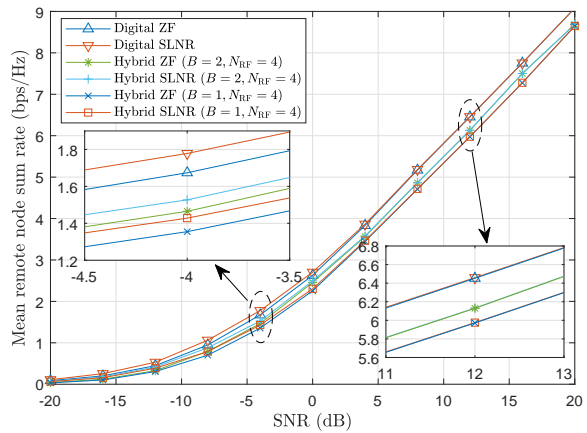}    
\caption{Mean remote node sum rate comparison of digital/hybrid SLNR/ZF beamforming schemes ($N_{\mathrm{T}}=8, K=3, N_{\mathrm{R},l}=1,\forall l$).}
\label{fig_Capacity_comparison}
\end{figure}

Mean remote node sum rates of hybrid SLNR beamforming in an independently and identically distributed (i.i.d.) Rayleigh fading channel were depicted in Fig. \ref{fig_Capacity_comparison}. Digital SLNR, BD ZF, and hybrid BD ZF were shown as reference. In general, SLNR slightly outperformed BD ZF in the low SNR regime. Performance of hybrid SLNR with 1 or 2 bits phase resolution analog precoder was approximately 15\% to 7\% worse than digital SLNR in the SNR range between -12 dB to 0 dB. This performance gap reduced to 5\% in the moderate SNR regime. Considering that the implementation of hybrid SLNR was significantly simpler than digital SLNR, the hybrid SLNR structure was able to provide better tradeoff between complexity and performance.

%is compared with ZF beamforming and digital SLNR beamforming in Fig. \ref{fig_Capacity_comparison}. 
%
%In the high SNR regime (SNR $\geq$ 10 dB), digital ZF beamforming achieves the highest sum rate when the dimensional condition is satisfied ($N_\mathrm{T}>N_\mathrm{R}(K-1)$), because interference from other users is eliminated. However, when the number of RF chain is small and the dimensional condition does not hold ($N_\mathrm{RF}<N_\mathrm{R}(K-1)$), the proposed hybrid SLNR beamforming significantly outperforms hybrid ZF beamforming. Performance loss can be observed in general when switching from digital SLNR to hybrid SLNR beamforming, due to the restriction on the analog precoder with finite quantization bits.

\begin{figure}[t]
\centering
\includegraphics[width=3.5in]{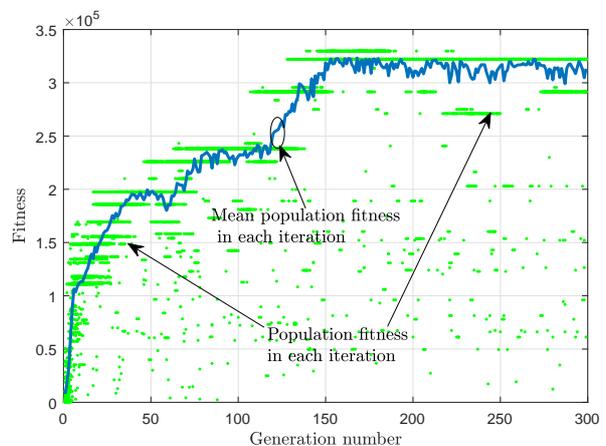}    
\caption{Fitness of population in terms of the number of generations ($N_{\mathrm{T}}=8, K=3, N_{\mathrm{R},l}=1,\forall l$, $B$=1, SNR=$10$dB)}
\label{fig_Fitnesslog_N8_K3_M2_SNR10dB}
\end{figure}

It is also important to study the complexity of the analog precoder design algorithm. Fig. \ref{fig_Fitnesslog_N8_K3_M2_SNR10dB} depicted the population fitness and mean fitness in terms of the number of generations. As the number of generation progressed, the fitness of the whole population tended to improve. Significant increase in fitness can be observed after a few generations. The improvement saturated when the number of generations reached approximately $150$. Hence, balance between performance and the number of generations should be reached and a reasonable number of iterations would be between 150 to $200$.  

\begin{figure}[h]
\centering
\subfloat[Subfigure 1 list of figures text][Digital SLNR beamforming]{
\includegraphics[width=3.5in]{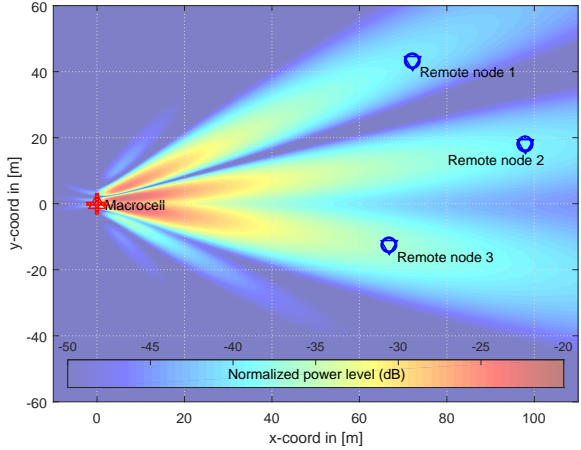}
\label{fig:subfig1}}
\qquad
\subfloat[Subfigure 2 list of figures text][Hybrid SLNR beamforming with $1$-bit phase shifters]{
\includegraphics[width=3.5in]{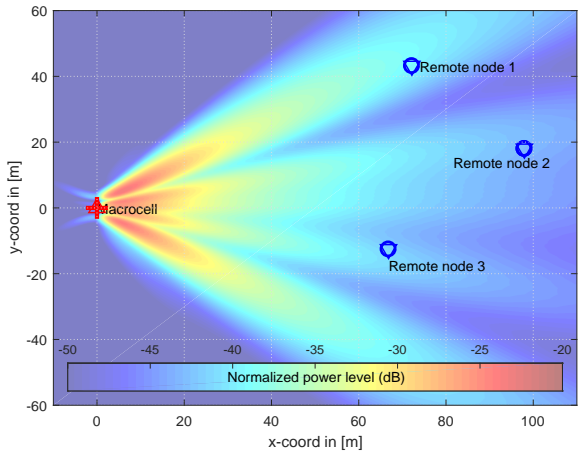}
\label{fig:subfig2}}

\caption{Beam patterns of digital and hybrid SLNR beamforming.}
\label{fig:globfig}
\end{figure}

Beam patterns of digital and hybrid SLNR beamforming were illustrated in Fig. \ref{fig:globfig}. It was assumed in this figure that remote nodes were in line-of-sight positions. It can be seen in Fig. \ref{fig:subfig1} that with digital SLNR beamforming, the shaped beams were sharp and with clear spatial boundaries. The hybrid SLNR beamforming with 1-bit phase shifters in Fig. \ref{fig:subfig2} had reasonably well shaped beams. However, remote nodes received weaker energy due to the limited capability of analog precoder. More importantly, a strong beam was observed in hybrid SLNR, which can cause intercell interference in a multicell network. This implied that current focus in the literature of hybrid beamforming design approximating the performance of fully digital beamforming were not sufficient. Impact of hybrid beamforming on multicell networks are essential as well. 

%%%%%%%%%%%Conclusion begins%%%%%%%%%%%%
\section{Conclusions} \label{conclusion_section}
A wireless fronthaul system with hybrid SLNR beamforming and a finite phase resolution analog precoder has been proposed in this paper. The digital precoder has been derived in closed-form while the analog precoder has been obtained using genetic algorithm. The design of analog precoder did not depend on the \textit{true} channel matrix, which was commonly assumed in the literature. Fronthaul links are relative stable providing quasi-static channels. Simulations have shown that the proposed hybrid SLNR beamforming was able to achieve similar performance to digital SLNR even with 1 or 2 bits of phase resolution in the analog precoder. It was also shown that hybrid beamforming could produce undesirable beams due to limited capability at the analog precoder. Therefore, designing algorithms for hybrid beamforming to approximate full digital beamforming in single cell scenario may not be sufficient. Impact of hybrid beamforming on multicell networks should be studied in future work.
%%%%%%%%%%%Conclusion ends%%%%%%%%%%%%%%
%\vspace{-0.3cm}
\section*{Acknowledgment}
Part of this work has been performed in the framework of the Horizon 2020 project ONE5G (ICT-760809) receiving funds from the European Union. The authors would like to acknowledge the contributions of their colleagues in the project, although the views expressed in this contribution are those of the authors and do not necessarily represent the project.

%%%%%%%%%%%%%%%%%%%%%%%%%%%%%%%%%%%%%%%%%%%%%%%%%%%%%%%%%%%%%%%%%%%%%%%%%%%%%%%%%%%%%%%%%%%%%%%%%
%%%%%%%%%%%%%%%%%%%%%%%%%%%%%%%%%%%%%%%%%%%%%%%%%%%%%%%%%%%%%%%%%%%%%%%%%%%%%%%%%%%%%%%%%%%%%%%%

%%%%%%%%%%%Bibliography ends%%%%%%%%%%%%%%

%\begin{biography}[{\includegraphics[width=1in,height=1.25in,clip,keepaspectratio]{mshell}}]{Michael Shell}


%\vfill


\end{document}